\documentclass[preprint,eqsecnum,showpacs,draft,pra,aps]{revtex4}

\begin{document}
\title{Optimal purifications and fidelity for
displaced thermal states}
\author{Paulina Marian}
\affiliation{ Department of Chemistry, University of 
Bucharest, Boulevard Regina Elisabeta 4-12, R-030018 Bucharest, Romania}
\author{Tudor A. Marian}
\affiliation{Department of Physics, University of Bucharest, P.O.Box MG-11,
R-077125 Bucharest-M\u{a}gurele, Romania}

\date{\today}

\begin{abstract}
  
 We evaluate the Uhlmann fidelity
between two one-mode displaced thermal  states as the maximal probability transition between  appropriate
 purifications of the  given states. The optimal  purifications defining  the 
fidelity  are proved  to be two-mode displaced Gaussian states.
\end{abstract}
\maketitle
\section{Introduction}
An important issue in quantum information theory is the
ability to distinguish between different quantum states, either
pure or mixed. Suppose 
we first deal with pure quantum states described
by the 
state vectors $|\Psi_1 \rangle$ and  $|\Psi_2 \rangle$.  Geometrically, a natural measure of their distinguishability is the angle between $|\Psi_1 \rangle$ and  $|\Psi_2 \rangle$ or any simple function of this angle.  Let us thus  define the {\em fidelity}  of the two pure states as the 
quantum-mechanical transition probability  $|\langle \Psi_1|
\Psi_2 \rangle|^2$. The so defined fidelity is a measure of the "closeness" of two pure quantum states. It is 1 when the two states coincide and 0 when they are orthogonal.

If one of the states, say 2, is mixed  we can still define
fidelity on quantum-mechanical grounds as
\begin{eqnarray}
{\cal F}(\rho_1,\rho_2)=\langle \Psi_1|\rho_2|\Psi_1\rangle. \label{1}
\end{eqnarray}
In Eq.\ (\ref{1}),   $\rho_1=|\Psi_1\rangle\langle \Psi_1|$ and $\rho_2$ are the 
density operators of the two quantum states acting on a Hilbert space ${\cal H}_{A}$.
Both situations (pure states or at least a pure state) may be described by
the formula
\begin{eqnarray}
{\cal F}(\rho_1,\rho_2)=Tr(\rho_1 \rho_2). \label{2} \end{eqnarray}
 However, when the two states 
are mixed, Eq.\ (\ref{2}) is unsatisfactory as  a probability transition and, 
consequently, as a  fidelity. Indeed, by applying Eq.\ (\ref{2}) to the case of mixed 
$\rho_1\equiv\rho_2$ we get ${\cal F}(\rho_1,\rho_1)=Tr(\rho_1^2) <1$, which  contradicts the natural expectation that  the fidelity of a state with itself is equal to 1. 
A new quantum-mechanical concept is thus necessary in order to define
a fidelity between mixed states. The usefulness of such a concept is better seen in communication theory where
 one has to evaluate quantitative measures of the faithfulness the mixed states are
transmitted within a quantum channel \cite{Sch}.  As a recent example,
in the continuous-variable teleportation \cite{br1} the accuracy of the process is usually measured  by the fidelity between the input and teleported states. 

  In Ref.\cite{Uhl}, Uhlmann introduced the fidelity between
the mixed states  $\rho_1$ and $\rho_2$
as the maximal quantum-mechanical transition probability 
between {\em all} purifications $|\Psi_{\rho_1}\rangle$ 
and $|\Psi_{\rho_2}\rangle$ of the given states:
\begin{equation}
{\cal F}(\rho_1, \rho_2):=\text{max}|\langle \Psi_{\rho_1}|\Psi_{\rho_2} 
\rangle|^2. \label{def} 
\end{equation}
The pure states $|\Psi_{\rho_1}\rangle$ and $|\Psi_{\rho_2}\rangle$
are defined in an extended Hilbert space ${\cal H}_{A}\otimes 
{\cal H}_{B}$ so that the given states are their reductions over the 
ancillary Hilbert space ${\cal H}_{B}$:
$\rho_1={\rm Tr}_{B} (|\Psi_{\rho_1}\rangle \langle\Psi_{\rho_1}|)$  and
$\rho_2={\rm Tr}_{B} (|\Psi_{\rho_2}\rangle \langle\Psi_{\rho_2}|)$.

Also, Uhlmann 
has found that the distance $D_{B}$ between two 
density operators discovered by Bures
\cite{Bur} is related to the transition probability ${\cal F}$
via 
\begin{equation}
D^2_{B}(\rho_1,\rho_2)=2[1-\sqrt{{\cal F}(\rho_1,\rho_2)}].
\label{db}\end{equation}
The explicit formula for the Bures-Uhlmann transition probability is 
\begin{eqnarray}
{\cal F}(\rho_1,\rho_2)=\left\{Tr[(\sqrt{\rho_1}\rho_2\sqrt{\rho_1})^{1/2}
]\right\}^2,\label{3} \end{eqnarray}
which simplifies to Eq.\ (\ref{2}) when  $\rho_1$ describes a pure state.

In the framework of estimation theory, Braunstein and Caves \cite{br} 
have defined a Riemannian metric
 on the space of density operators by generalizing the notion of statistical 
distance introduced by Wootters for pure states \cite{Woo}. This metric
was derived by imposing the optimal statistical distinguishability between 
neighboring quantum states. The statistical distance coincides, up to a
$1/2$ factor, with 
the Bures distance \ (\ref{db}). Therefore the transition probability
\ (\ref{3}) has a sound interpretation in the geometry of mixed
quantum states \cite{fuchs,Jo}.

Explicit expressions for the fidelity of mixed quantum
states were first given by using Eq.\ (\ref{3}) for finite dimensional
cases. In two and three dimensions, the problem was solved by H\"ubner in 
Refs.\cite{hub}.
Evaluation of the fidelity in the infinite dimensional
Hilbert space of the density operators is now 
of greatest importance
due to the experimental interest in quantum information processing 
of  field states \cite{br1}. Results in the continuous-variable case are only available for one- and two-mode Gausssian states.
Especially useful in experiments, the Gaussian states are characterized by the exponential form of their density operators. All the explicit results concerning the fidelity between single-mode 
  Gaussian states \cite{Tw,Sc1} were
obtained by exploiting the Bures-Uhlmann formula,
 Eq.\ (\ref{3}).  In recent years the Bures distance proved to be a reliable tool in quantifying nonclassicality (in the one-mode case \cite{PTHlet}) and inseparability (for two-mode states \cite{PTH1}). 
 The explicit formula of the fidelity was then used to quantify the
 faithfulness of  teleportation for mixed  one-mode  Gaussian states 
through a Gaussian channel in  Refs.\cite{PTH03,ban}.
 
In the present paper we take advantage of previous  results on fidelity in the one-mode Gaussian case and prove that the optimal purification in defining  fidelity via Eq.\ (\ref{def}) is a two-mode Gaussian state. We apply the maximization procedure in Eq.\ (\ref{def}) by conjecturing on the structure of the required purifications.   Our point here is to prove that by this conjecture we recover the formula for one-mode fidelity given in Ref.\cite{Sc1}.   We choose to deal with the simplest displaced Gaussian states, namely the thermal coherent states 
(TCS's) in order to have just a one-parameter maximization problem. Our aim here is to present the novelty of the method rather than complicated analytic evaluations \cite{Mar3}. 
In Sec. II we recall the eigenvalue problem of the thermal
density operator, write the general expansion for a Schmidt purification of TCS's and evaluate its characteristic function (CF). In deriving the fidelity between two TCS's in Sec. III  we follow the principal line of reasoning given by Jozsa 
in the very clear paper \cite{Jo}.
Although his derivation is valid for a finite-dimensional Hilbert
space of the density operators, we explicitly show that it may be 
extended to the case of  Gaussian states. Our conclusions are drawn in Sec. IV.

\section{Purifications of a TCS}

\subsection{Eigenvalue-problem for the thermal density operator}

 The thermal density operator,
\begin{equation}
\rho_T=\frac{1}{\bar{n}+1}\sum_{m=0}^{\infty} \left(\frac{\bar{n}}{
\bar{n}+1}\right)^m|m \rangle \langle m|,\label{4}\end{equation}
where 
\begin{equation}
\bar{n}=\left[\exp{\left(\frac{\hbar \omega}{K_BT}\right)}-1\right]^{-1}
\label{temp}\end{equation}
is the mean occupancy at the temperature $T$,
has a discrete, nondegenerate and positive spectrum
of eigenvalues 
\begin{equation}
\eta_{j}=\frac{1}{\bar{n}+1} s^j,\;\;s:=\left(\frac{\bar{n}}{
\bar{n}+1}\right),\;\;j=0,1,2...\label{eigv}\end{equation}
The corresponding orthonormal eigenvectors form the Fock basis $\{|j \rangle \}$ in the
space. We thus have 
\begin{equation}
\rho_T|j \rangle=\eta_j|j \rangle.\label{eig}\end{equation}
Subsequent unitary actions $U$ on the thermal density operator, $\rho_T \rightarrow  
 \rho=U\rho_T U^{\dag}$, do not modify the spectrum of eigenvalues while
the corresponding orthonormal system becomes $
\{U|j\rangle\}$.

An example of unitary action on $\rho_T$ is the displacement realized 
by the Weyl operator $D(\alpha)=\exp{(\alpha a^{\dag}-\alpha^* a)}$
($a$ denotes the annihilation operator),

\begin{equation}
\rho_{TC}:=D(\alpha)\rho_{T}D^{\dag}(\alpha).\label{tcs}\end{equation}
 
Equation \ (\ref{tcs})  describes a
TCS, namely a coherent state 
with thermal noise. A TCS is 
a classical mixed Gaussian state, having a well behaved $P$-representation for any values of the parameters $\alpha$ and $T$. 

\subsection{Building purifications of a TCS}
 We use the Schmidt polar form \cite{Jo,Hu} to build purifications
lying in an extended (two-mode) Hilbert space 
${\cal H}_e={\cal H}\otimes {\cal H}$,
where ${\cal H}$ is the (single-mode) Hilbert space 
 of the TCS's. Let $|\Phi\rangle$ be a 
purification of $\rho_{TC}$, Eq.\ (\ref{tcs}),
\begin{equation}
|\Phi \rangle=\sum_{n=0}^{\infty} \sqrt{\eta_n} D_1(\alpha)|n \rangle\otimes
D_2(\beta)|n \rangle.\label{p}\end{equation}
$|\Phi\rangle$ has
the reduced mode 1 in the TCS with $\rho_{TC}=D_1(\alpha)\rho_{T}D_1^{\dag}(\alpha)$  and 
the reduced mode 2 in the TCS having the same temperature $T$ but 
a different displacement $\beta$. It is interesting to write down its characteristic function  $\chi(\lambda_1,\lambda_2)$ defined as the expectation value of the displacement operator $D(\lambda_1,\lambda_2)=D_1(\lambda_1) D_2(\lambda_2)$. We have
\begin{equation}\chi(\lambda_1,\lambda_2)=\langle\Phi|D(\lambda_1,\lambda_2)|\Phi \rangle.\label{cf}\end{equation}
By using  Eq.\ (\ref{p}), the CF \ (\ref{cf}) becomes
\begin{equation}
\chi(\lambda_1,\lambda_2)=\sum_{n,m=0}^{\infty} \sqrt{\eta_n\eta_m}
\langle m|D_1^{-1}(\alpha)D_1(\lambda_1) D_1(\alpha)|n\rangle
\langle m|D_2^{-1}(\beta) D_2(\lambda_2)D_2(\beta)  |n\rangle. \label{cf1}\end{equation}
In the above equations the subscripts $1$ and $2$ denote the two reduced modes.
 We need now to recall several properties of the displacement operators 
\cite{Gl}:
\begin{enumerate}
\item
 the multiplication law of the Heisenberg-Weyl group
\begin{equation} D(\alpha)D(\beta)=
\exp{[\frac{1}{2}(\alpha\beta^{\ast}-\alpha^{*}\beta})]\;D(\alpha+\beta).
\label{7a}\end{equation} 
 \item the matrix elements in the Fock basis
\begin{eqnarray}
<k|D(\beta)|l>=\left(\frac{l!}{k!}\right)^{1/2} \beta^{k-l}
\exp{(-\frac{|\beta|^2}{2})} L_{l}^{(k-l)}(|\beta|^2),
\label{7}\end{eqnarray}
where $L_{l}^{(k-l)}$ is a Laguerre polynomial.
\end{enumerate}
Application of these properties to Eq.\ (\ref{cf1}) leads us to the double summation
\begin{eqnarray}
\chi(\lambda_1,\lambda_2)&=&\exp{\left[-\frac{|\lambda_1|^2}{2}-\frac{|\lambda_2|^2}{2}+\lambda_1\alpha^*-\lambda_1^*\alpha  +\lambda_2\beta^*-\lambda_2^*\beta\right]}\nonumber\\ &&  \times \frac{1}{
\bar{n}+1}\sum_{n,m=0}^{\infty}\frac{n!}{m!}s^{(m+n)/2}
(\lambda_1\lambda_2)^{m-n}L_{n}^{(m-n)}(|\lambda_1|^2)
L_{n}^{(m-n)}(|\lambda_2|^2).\label{cf2}\end{eqnarray}
The summation with respect to $n$ is carried out by employing
an important bilinear series involving Laguerre polynomials \cite{Bu}.
 As a result of the summation with respect to $m$, we are then left with the  generating function of the Laguerre
polynomials \cite{Bu}.
Finally the CF of the purification $|\Phi\rangle$ reads 
\begin{eqnarray}
\chi(\lambda_1,\lambda_2)&=&\exp{\left[-(\bar{n}+\frac{1}{2})(|\lambda_1|^2+|\lambda_2|^2)+\sqrt{\bar{n}(\bar{n}+1)}(\lambda_1\lambda_2+\lambda_1^*\lambda_2^*)\right]}\nonumber\\ 
 &&\times \exp{\left[\lambda_1\alpha^*-\lambda_1^*\alpha  +\lambda_2\beta^*-\lambda_2^*\beta\right]}.
\label{cf3}\end{eqnarray}
Equation \ (\ref{cf3}) is the CF of a two-mode Gaussian state having the displacement parameters  
$\alpha $  (mode 1) and $\beta$ (mode 2), respectively. The covariance matrix of this state can be easily written by examining the undisplaced part of its CF (first line of the right-hand side of Eq.\ (\ref{cf3})). We get
\begin{eqnarray}{\cal V}=\left(\begin{array}{cccc}\bar{n}+\frac{1}{2} &0& \sqrt{\bar{n}(\bar{n}+1)} &0\\
0& \bar{n}+\frac{1}{2}&0&-\sqrt{\bar{n}(\bar{n}+1)}\\ \sqrt{\bar{n}(\bar{n}+1)}  &0&\bar{n}+\frac{1}{2}&0\\0
&-\sqrt{\bar{n}(\bar{n}+1)}&0&\bar{n}+\frac{1}{2} \end{array}\right).\label{cm}
\end{eqnarray}
It is easy to verify that the covariance matrix \ (\ref{cm}) describes a symmetric  state having $\det( {\cal V} )=1/16$ as expected for a pure two-mode Gaussian state.    
\section{Fidelity}
 We evaluate now the fidelity
between  two TCS's with the density operators
$\rho_{TC}^{(I)}=D(\alpha_1)\rho_{T_1}D^{\dag}(\alpha_1)$ and $\rho_{TC}^{(II)}=D(\alpha_2)\rho_{T_2}D^{\dag}(\alpha_2)$
 corresponding to the temperatures $T_1$ and $T_2$ and having the  displacement parameters 
$\alpha_1$ and $\alpha_2$,
respectively.   
Let $|\Phi^{(I)}\rangle$ be a specified  
purification of $\rho_{TC}^{(I)}$ defined by its CF
\begin{equation}
\chi^{(I)}(\lambda_1,\lambda_2)=\exp{\left[-(\bar{n}_1+\frac{1}{2})(|\lambda_1|^2+|\lambda_2|^2)+\sqrt{\bar{n}_1(\bar{n}_1+1)}(\lambda_1\lambda_2+\lambda_1^*\lambda_2^*)+\lambda_1\alpha_1^*-\lambda_1^*\alpha_1  \right]}.
\label{p1}\end{equation}
According to  Eq.\ (\ref{cf3}), the two-mode Gaussian state
 $|\Phi^{(I)}\rangle$ has
the reduced mode 1 in the TCS $\rho_{TC}^{(I)}$  and 
the reduced mode 2 in a thermal state at the same $T_1$ temperature. To handle Eq.\ (\ref{def}), we  keep fixed the purification $|\Phi^{(I)}\rangle$ and show that there exists a purification $|\tilde{\Phi}^{(II)}\rangle$ of the state $\rho_{TC}^{(II)}$ which realizes the maximimum of the transition probability such that
\begin{equation}|\langle \Phi^{(I)}|\tilde{\Phi}^{(II)}\rangle|^2={\cal F}(\rho_{TC}^{(I)},\rho_{TC}^{(II)}).\label{p11}\end{equation}
Let us now conjecture that the optimal purification $|\tilde{\Phi}^{(II)}\rangle$ belongs to the set of states $\{|\Phi^{(II)}\rangle\}$  having the CF of the general type, Eq.\ (\ref{cf3}):
\begin{eqnarray}
\chi^{(II)}(\lambda_1,\lambda_2)&=&\exp{\left[-(\bar{n}_2+\frac{1}{2})(|\lambda_1|^2+|\lambda_2|^2)+\sqrt{\bar{n}_2(\bar{n}_2+1)}(\lambda_1\lambda_2+\lambda_1^*\lambda_2^*)\right]}\nonumber \\&& \times \exp{\left[\lambda_1\alpha_2^*-\lambda_1^*\alpha_2  +\lambda_2\beta^*-\lambda_2^*\beta\right]}.
\label{p2}\end{eqnarray}
 This  
purification has the reduced mode 1 in the state
$\rho_{TC}^{(II)}$,  while the 
density operator of the second reduced system describes a TCS as well: $\rho^{\prime}_{TC}=
D(\beta)\rho_{T_2}D^{\dag}(\beta)$. 

 The transition probability between the purifications 
$|\Phi^{(I)}\rangle$ and $|\Phi^{(II)}\rangle$ can be found  by using their CF's, Eqs.\ (\ref{p1}) and \ (\ref{p2}): 
\begin{equation}
|\langle \Phi^{(I)}|\Phi^{(II)}\rangle|^2=\frac{1}{\pi^2}\int {\rm d}^2 \lambda_1
{\rm d}^2 \lambda_2\chi^{(I)}(\lambda_1,\lambda_2)
\left(\chi^{(II)}(\lambda_1,\lambda_2)\right)^*.
\label{ps}\end{equation}
This is a Gaussian integral of the type solved in the Appendix A of Ref.\cite{Mar1}. We easily obtain the transition probability
\begin{eqnarray}
|\langle\Phi^{(I)}|\Phi^{(II)}\rangle |^2&=&\frac{1}{[\sqrt{
(\bar{n}_1+1)(\bar{n}_2+1)}-\sqrt{\bar{n}_1 \bar{n}_2}]^2}\nonumber\\&&
\times
\exp{\left\{-\frac{1+\sqrt{s_1s_2}}{1-\sqrt{s_1s_2}}[|\beta|^2+|\alpha_1
-\alpha_2|^2]\right\}}\nonumber\\&&
\times \exp{\left\{\frac{\sqrt{s_1}+\sqrt{s_2}}{1-\sqrt{s_1s_2}}
[\beta(\alpha_2-\alpha_1)+\beta^*(\alpha^*_2-\alpha^*_1)]\right\}}.
\label{8}\end{eqnarray}
In Eq.\ (\ref{8}) we used the notations $s_1=\bar{n}_1/(\bar{n}_1+1)$ and $ s_2=\bar{n}_2/(\bar{n}_2+1)$.
 Now  we have to maximize \ (\ref{8}) with respect to the coherent 
amplitude $\beta$. An elementary calculation gives us the 
displacement parameter $\tilde{\beta}$ that realizes the maximum of the transition probability 
\begin{equation}
\tilde{\beta}=\frac{\sqrt{s_1}+\sqrt{s_2}}{1+\sqrt{s_1s_2}}
(\alpha^*_2-\alpha^*_1).\label{max1}\end{equation}
Interestingly, the optimal coherent amplitude $\tilde{\beta}$ depends on both
 temperatures in a nontrivial way. When $T_1=T_2$, $\tilde{\beta}$
is still temperature-dependent.   
By inserting $\tilde{\beta}$ in Eq.\ (\ref{8}) we get the 
Bures-Uhlmann transition probability (fidelity) between the
mixed states $\rho_{TC}^{(I)}$ and $\rho_{TC}^{(II)}$
\begin{equation}
{\cal F}(\rho_{TC}^{(I)},\rho_{TC}^{(II)})=F(\rho_{T_1},\rho_{T_2})
\exp{\left\{-\frac{|\alpha_1-\alpha_2|^2}{\bar{n}_1+\bar{n}_2+1}
\right\}},\label{fid1}\end{equation}
where
\begin{equation}
{\cal F}(\rho_{T_1},\rho_{T_2}):=\frac{1}{[\sqrt{
(\bar{n}_1+1)(\bar{n}_2+1)}-\sqrt{\bar{n}_1 \bar{n}_2}]^2},
\label{fid2}\end{equation}
is the fidelity between the thermal states corresponding to the
temperatures $T_1$ and $T_2$.
Equation \ (\ref{fid1}) is in agreement with the previous results 
of P\u{a}r\u{a}oanu and Scutaru \cite{Sc1}. 
In this way we have determined not only the fidelity but also 
{\it the most parallel pure entangled states which satisfy the requirement
of having the mixed states  $\rho_{TC}^{(I)}$ and
$\rho_{TC}^{(II)}$ as reduced states}.

\section{Conclusions}
To conclude, in this paper we have given an example of evaluation
 of the transition probability between two mixed Gaussian states
by applying explicitly the concepts of Bures and Uhlmann about
the distance between density operators. 
We succeeded to determine the  most parallel two-mode states having the
single-mode subsystems described by $\rho_{TC}^{(I)}$ and 
$\rho_{TC}^{(II)}$. 
The principal result of our paper is stated here: finding a full agreement between the expression of the fidelity obtained via Eq.\ (\ref{3}) \cite{Sc1}, and via Eq.\ (\ref{def}) (present work)  it follows that the optimal purifications in defining the probability of transition between mixed Gaussian states are also Gaussian \cite{Mar3}.
It is also remarkable that, by avoiding to apply
the general formula \ (\ref{3}), we  have reached the result
\ (\ref{fid1}) by straightforward elementary analytic means. 
\section*{Acknowledgement}
This work was supported by the Romanian 
MEC 
through the grant  CEEX 05-D11-68/2005 for the University of Bucharest.


\begin{thebibliography}
{100}
\bibitem{Sch}B. Schumacher, Phys. Rev. A {\bf 51}, 2738 (1995).
\bibitem{br1} S. L. Braunstein and H. J. Kimble, Phys. Rev. Lett. {\bf 80},
 869 (1998);  
 A. Furusawa {\it et al.}, Science {\bf 282}, 706 (1998).
\bibitem{Uhl}A. Uhlmann, Rep. Math. Phys. {\bf 9}, 273 (1976);
 Rep. Math. Phys. {\bf 24}, 229 (1986);
\bibitem{Bur} D. Bures, Trans. Am. Math. Soc. {\bf 135}, 199 (1969).
\bibitem{br} S. L. Braunstein and C. Caves, Phys. Rev. Lett. {\bf 72},
3439 (1994).
\bibitem{Woo} W. K. Wootters, Phys. Rev. D. {\bf 23}, 357 (1981).
\bibitem{fuchs} C. A. Fuchs, Ph. D. thesis, University of New Mexico, 1995, 
(quant-ph/9601020/1996).
\bibitem{Jo}R. Jozsa, J. Mod. Opt. {\bf 41}, 2315 (1994).
\bibitem{hub}M. H\"ubner, Phys. Lett. A {\bf 163}, 239 (1992);
M. H\"ubner, Phys. Lett. A {\bf 179}, 226 (1993).
\bibitem{Tw} J. Twamley, J. Phys. A: Math. Gen. {\bf 29}, 3723 (1996).
\bibitem{Sc1} Gh.-S P\u{a}r\u{a}oanu and H. Scutaru, Phys. Rev. A 
{\bf 58}, 869 (1998); H. Scutaru, J. Phys. A: Math. Gen. {\bf 31}, 3659 (1998).
\bibitem{PTHlet} Paulina Marian, T.A. Marian, and H. Scutaru,
Phys. Rev. Lett. {\bf 88}, 153601 (2002);  Paulina Marian, T.A. Marian, 
and H. Scutaru,
Phys. Rev. A. {\bf 69}, 022104 (2004).
\bibitem{PTH1} Paulina Marian, T. A. Marian, 
and H. Scutaru, Phys. Rev. A {\bf 68},
 062309 (2003); M. C. de Oliveira, Phys. Rev. A {\bf 72},
 012317 (2005); Paulina Marian and  T. A. Marian, e-print quant-ph/0705.1138.
\bibitem{PTH03}Paulina Marian, T. A. Marian, and H. Scutaru, Rom. J. Phys.
{\bf 48}, 727 (2003)(e-print quant-ph/0601045). 
\bibitem{ban}M. Ban, Phys. Rev. A {\bf 69}, 054304 (2004).
 \bibitem{Mar3}    The method presented in the body of the paper 
 can be easily extended to
arbitrary single mode Gaussian states. The volume of analytical work 
is however larger in the general case. The conclusion is the same as in the present work: the optimal purifications are two-mode Gaussian states.
\bibitem{Hu} L. Hughston, R. Jozsa, and W. Wootters, Phys. Lett. A
{\bf 183}, 14 (1993).
\bibitem{Gl}  K. E. Cahill and R. J. Glauber  Phys. Rev.
 {\bf 177}, 1882  (1969).
\bibitem{Bu}H. Buchholz, 1969 {\it The Confluent Hypergeometric Function}
Springer, Berlin, 1969. See Eq.(5), p. 152, for the bilinear series and 
Eq.(11a), p. 138, for the generating function of the Laguerre polynomials.
\bibitem{Mar1} Paulina Marian and T. A. Marian, Phys. Rev. A {\bf 47},
4474 (1993).

\end{thebibliography}
\end{document}